\documentclass[twocolumn,showpacs,amsmath,amssymb,pre]{revtex4}

\usepackage{graphicx}
\usepackage{dcolumn}
\usepackage{bm}
\usepackage{epsf}
\usepackage{amsmath}
\usepackage{amssymb}
\usepackage{color}

\newcommand{\la}{\left\langle}
\newcommand{\ra}{\right\rangle}

\newcommand{\bla}{bla\\bla\\bla\\bla\\bla}
\newcommand{\PR}{Phys. Rev. }

\newcommand{\PRE}{Phys. Rev. E }
\newcommand{\PRL}{Phys. Rev. Lett. }
\newcommand{\EPL}{Europhys. Lett. }

\begin{document}

\title{
Optimal performance of a quantum Otto refrigerator}

\author{Obinna Abah}

\affiliation{Department of Physics, Friedrich-Alexander-Universit\"at  Erlangen-N\"urnberg, D-91058 Erlangen, Germany}

\author{Eric Lutz}
\affiliation{Department of Physics, Friedrich-Alexander-Universit\"at  Erlangen-N\"urnberg, D-91058 Erlangen, Germany}


\begin{abstract} 
We consider a quantum Otto refrigerator  cycle of a time-dependent harmonic oscillator. We investigate the coefficient of performance at maximum figure of merit for adiabatic and nonadiabatic frequency modulations. We obtain analytical expressions for the optimal performance both in the high-temperature (classical) regime and in the low-temperature (quantum) limit. We moreover analyze the breakdown of the cooling cycle for strongly nonadiabatic driving protocols and derive analytical estimates for the minimal driving time allowed for cooling. 
\end{abstract}

\maketitle

\section{Introduction}

Heat engines and refrigerators are two prominent examples of thermal machines. While heat engines  produce work by absorbing heat from a hot reservoir, refrigerators  consume work to extract heat from a cold reservoir \cite{cal85}. In a sense, refrigerators  thus appear as heat engines functioning in reverse. In  equilibrium thermodynamics, this symmetry  is reflected in  their maximum efficiency (traditionally called 'coefficient of performance' for refrigerators \cite{cal85}): The maximum efficiency of any heat engine operating cyclically between two thermal reservoirs at temperatures  $T_1$ and $T_2$ ($T_1 < T_2$)  is given by the Carnot formula, $\eta_\mathrm{c} = \text{(work output)}/\text{(heat input)} = (T_2 -T_1)/T_2$. On the other hand, the corresponding maximum coefficient of performance of any refrigerator is $\epsilon_\mathrm{c} = \text{(heat input)}/\text{(work input)}  =T_1/(T_2 - T_1)$ \cite{cal85}.
However,  this analogy between the two devices is only superficial and their finite-time behavior is radically different.

The theory of equilibrium thermodynamics has lately been extended in two different directions: finite-time thermodynamics \cite{and84,and11} and quantum thermodynamics \cite{gem09,lev14}. The equilibrium Carnot expressions  $\eta_\mathrm{c}$ and  $\epsilon_\mathrm{c}$ are only of limited practical importance since they may only be reached  for machines that run infinitely slowly. In this quasistatic limit,  a core assumption of equilibrium thermodynamics  \cite{cal85}, the power output of a device vanishes. The efficiency at maximum power $\eta^*$ of a heat engine, and the corresponding coefficient of performance at maximum figure of merit $\epsilon^*$ of a refrigerator, are two quantities of far greater significance for real machines that operate in finite time. Their investigation via the optimization of nonequilibrium  processes beyond the quasistatic approximation is one of the main goals of finite-time thermodynamics \cite{and84,and11}. A central result for the efficiency at maximum power of a heat engine is the so-called Curzon-Ahlborn efficiency, $\eta_\text{ca} = 1 - \sqrt{1 - \eta_\mathrm{c}}$, obtained independently by Reitlinger \cite{rei29}, Chambadal \cite{cha49}, Yvon \cite{yvo55}, Novikov \cite{nov57} and Curzon-Ahlborn \cite{cur75} (see Refs. \cite{vau14,oue14} for a history of the formula). At about the same time, starting with the work of Scovil and Schulz-Dubois on maser heat engines \cite{sco59}, refrigerators \cite{geu59} and heat pumps \cite{geu67}, the investigation of thermal machines has been extended to the low-temperature, quantum regime \cite{ali79,kos84,he02,kie04,ton05,all08,rez06}. Interestingly, a variety of theoretical studies have shown that the performance of quantum thermal machines may be  enhanced beyond the classical Carnot limit  \cite{scu01,scu02,scu03,dil09,man13,hua12,ros14,aba14,cor14}. A standard model of a quantum heat engine consists of a single harmonic oscillator \cite{kos84,he02} (or equivalently a two-level system \cite{fel00,qua07}) that is alternately connected to a hot and a cold reservoir. Two  important cases are usually distinguished  \cite{rez06}: the  adiabatic limit which corresponds to slow expansion/compression phases (compared to the free dynamics of the system) and the sudden limit of fast expansion/compression. In both instances, the finite-time performance of the machine can be studied analytically. A detailed investigation of the efficiency at maximum power of a quantum harmonic  heat engine in these two limits  may be found in Ref.~\cite{rez06} (and references therein) in the high-temperature (classical) regime. As expected, the efficiency at maximum power $\eta^*$ coincides with the classical Curzon-Ahlborn expression for adiabatic driving. However, $\eta^*$ is much lower for  sudden driving and  is found to be  bounded from above by one 
half instead of one. These results have been  generalized to the low-temperature (quantum) domain in Ref.~\cite{aba12} and were shown to depend explicitly on the reduced Planck constant $\hbar$ (see also Ref.~\cite{zha14}).

The finite-time analysis of the performance of refrigerators is more involved than that of heat engines. It has for this reason only been  completed much later \cite{yan84,yan90,vel97}. One of the main difficulties is the identification of a proper optimization criterion: the maximization of the cooling power or the minimization of the power input, for instance, do not result in a temperature-dependent bound    for the coefficient of performance, in contrast to what is known for heat engines \cite{vel97}. The finite-time counterpart of the Curzon-Ahlborn efficiency for refrigerators was first obtained by Yan \cite{yan84} and Yan and Chen \cite{yan90} in the late 1980s and rediscovered   by Velasco and coworkers in 1997 \cite{vel97} (see also Ref.~\cite{all10}). It is given by $\epsilon_\text{yan} = \sqrt{1 + \epsilon_{\mathrm{c}}} -1$ with an optimization criterion (the figure of merit) that is taken as the product  of the coefficient of performance  $\epsilon$  and the cooling power of the refrigerator (see eq.~(7) below). It has recently been  shown that this  optimization criterion for refrigerators is indeed in true correspondence to the maximum power criterion for heat engines \cite{tom12}. Meanwhile, additional results for the coefficient of performance at maximum figure of merit  have been reported for quantum refrigerator models \cite{bir08,yua14}. However, to our knowledge, only the high-temperature (classical) expression of $\epsilon^*$  in the adiabatic limit has been determined so far. Analytical results  are  missing for the important low-temperature (quantum) domain, both for adiabatic and fast driving, as well as for fast  driving in  the classical  regime.

In this paper, we consider a quantum Otto refrigerator cycle for a time-dependent harmonic oscillator, a paradigmatic model for a quantum thermal machine \cite{lin03,lin03a,san04,rez09,tor13}. We first derive an exact formula for its coefficient of performance that is valid for any frequency driving  and reservoir temperatures. We  use this equation to evaluate the optimal coefficient of performance both in the high-temperature (classical) and low-temperature (quantum) regime, for  slow and fast frequency modulations. We further show that the cooling cycle breaks down for strongly nonadiabatic driving and present an approximate expression for the minimal driving time permitted for cooling.

\section{Quantum Otto refrigerator}
\begin{figure}[t]
\includegraphics[width=\columnwidth]{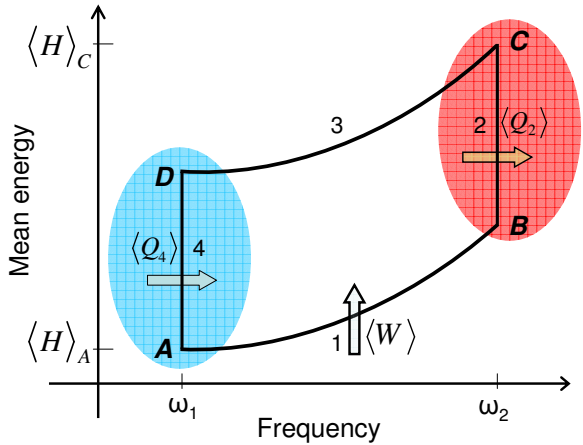}
\caption{Energy-frequency diagram of a quantum Otto refrigerator. The thermodynamic cycle consists of two isentropic  processes (strokes 1 and 3) and  two isochoric  processes (strokes 2 and 3). During one full cycle work is consumed by the refrigerator to pump heat from the cold to the hot reservoir. 
\label{cycle}}
\end{figure}

We study a quantum Otto refrigerator  based on a  harmonic oscillator with time-dependent frequency. The four branches of the cycle are (see Fig.~\ref{cycle}):
(1) Isentropic compression $A\rightarrow B$: the  oscillator is isolated from the reservoirs and its frequency  is increased from $\omega_1$ to $\omega_2$. Work is added to the system during this process whereas entropy is constant.
(2) Hot isochore $B \rightarrow C$: the frequency is kept fixed  while the oscillator thermalizes to state $C$ characterized by the inverse temperature $\beta_2= (k_B T_2)^{-1}$ of the hot reservoir. 
(3) Isentropic expansion $C \rightarrow D $: the frequency is decreased back to its initial value at constant entropy.
(4) Cold isochore $D \rightarrow A$: the system is brought back to the initial thermal state $A$ by coupling it, at fixed frequency, to the cold reservoir with inverse temperature $\beta_1= (k_B T_1)^{-1}$. 
Expansion/compression  phases need  not   be infinitely slow as usually assumed in equilibrium thermodynamics.

 The analysis  of the performance of the refrigerator requires the evaluation of the mean energies of the quantum oscillator at the four corners of the cycle, in analogy  to the case of the quantum Otto heat engine \cite{aba12}:
\begin{subequations}
\label{s1}
\begin{eqnarray}
\la H\ra_{A}&=& \frac{\hbar\omega_1}{2} \coth\left(\frac{\beta_1\hbar \omega_1}{2}\right),\\
\la H\ra_{B}&=& \frac{\hbar\omega_2}{2}\,Q^\ast_1 \coth\left(\frac{\beta_1 \hbar \omega_1}{2}\right),\\ 
\la H\ra_{C}&=& \frac{\hbar\omega_2}{2} \coth\left(\frac{\beta_2 \hbar\omega_2}{2}\right),\\ 
\la H\ra_{D}&=& \frac{\hbar\omega_2}{2} Q^\ast_2\coth\left(\frac{\beta_2 \hbar\omega_2}{2}\right). \end{eqnarray}
\end{subequations}
We have here introduced  the dimensionless quantity $Q^\ast_{1,2}$ which  depends on  the  speed of the frequency driving  \cite{hus53}. It is equal to $Q^\ast_{1,2}=1$ for quasistatic frequency modulation and to $Q^\ast_{1,2}= (\omega_1^2+\omega_2^2)/(2\omega_1\omega_2)$ for a sudden frequency switch.  The explicit expression of $Q^\ast_{1,2}$ for arbitrary frequency modulation can be found in Refs. \cite{def08,def10}. 

The work done on the system during the first and third strokes of the cycle is given by,
\begin{equation}
\la W_1\ra = \la H\ra_B - \la H\ra_A = \left(\frac{\hbar \omega_2}{2}Q^\ast_1 - \frac{\hbar \omega_1}{2}\right)\coth\left(\frac{\beta_1\hbar\omega_1}{2}\right)
\end{equation}
and 
\begin{equation}
\la W_3\ra = \la H\ra_D - \la H\ra_C = \left(\frac{\hbar\omega_1}{2}Q^\ast_2 - \frac{\hbar \omega_2}{2}\right)\coth\left(\frac{\beta_2\hbar\omega_2}{2}\right).
\end{equation}
On the other hand, the heat extracted during the fourth stroke, $\la Q_4\ra=  \la H\ra_A - \la H\ra_D$, (the cooling part) reads,
\begin{equation}
\la Q_4\ra = \frac{\hbar \omega_1}{2} \left[\coth\left(\frac{\beta_1\hbar\omega_1}{2}\right) - Q^\ast_2\coth\left(\frac{\beta_2\hbar\omega_2}{2}\right)\right].
\end{equation}
For a refrigerator, heat is absorbed from the cold reservoir, $\la Q_4\ra \ge 0$ and flows into the hot reservoir, $\la Q_2\ra \le 0$. The condition for cooling is thus that $\omega_2/\omega_1 > \beta_1/\beta_2$.

 The coefficient of performance $\epsilon$ of  the Otto refrigerator, defined as the ratio of the heat removed from the cold reservoir, $\la Q_4\ra$, to the total amount of work done per cycle, $\la W\ra = \la W_1\ra + \la W_3\ra$, follows  as,

\begin{eqnarray}
\epsilon &=& \frac{\la Q_4\ra}{\la W_1\ra + \la W_3\ra} \\
 &=& \frac{\omega_1[\text{c}(\beta_1\hbar\omega_1/2)- Q^\ast_2\text{c}(\beta_2\hbar\omega_2/2)]}{(\omega_2Q^\ast_1-\omega_1)\text{c}(\beta_1\hbar\omega_1/2) - (\omega_2 - \omega_1Q^\ast_2)\text{c}(\beta_2\hbar\omega_2/2)}, \nonumber
\end{eqnarray}
where we have defined the function $\text{c}(x) = \coth(x)$.
The above quantum expression is exact and valid for any frequency modulation.
In the high-temperature limit, $\beta_i \hbar\omega_i\ll 1\, (i = 1,2)$, and for adiabatic processes, $Q^\ast_{1,2}=1$,  the coefficient of performance  reduces to \cite{rez06,bir08,yua14}, 
\begin{equation}
\epsilon_{ad} = \frac{\omega_1}{\omega_2 - \omega_1}.
\end{equation}
The above equation is  positive provided that $\omega_2 > \omega_1$ and is always smaller  than the Carnot expression, $\epsilon<\epsilon_c $.

\section{Performance at maximum figure of merit}
In order to optimize the performance of  the quantum refrigerator, we introduce the figure of merit $\chi$ defined as  the product of the heat absorbed from  the cold reservoir $\la Q_4\ra$, eq.~(4), and the coefficient of performance $\epsilon$, eq.~(5), over the duration of a thermodynamic cycle $t_{cycle}$ \cite{yan84,yan90,vel97,all10,tom12},
\begin{equation}
\chi = \frac{\epsilon \la Q_4\ra}{t_{cycle}}.
\end{equation}
We next compute the coefficient of performance at maximum figure of merit $\epsilon^*$ for slow and fast frequency transformations, both in the classical and quantum limits.

\begin{figure}[t]
\includegraphics[width=.95\columnwidth]{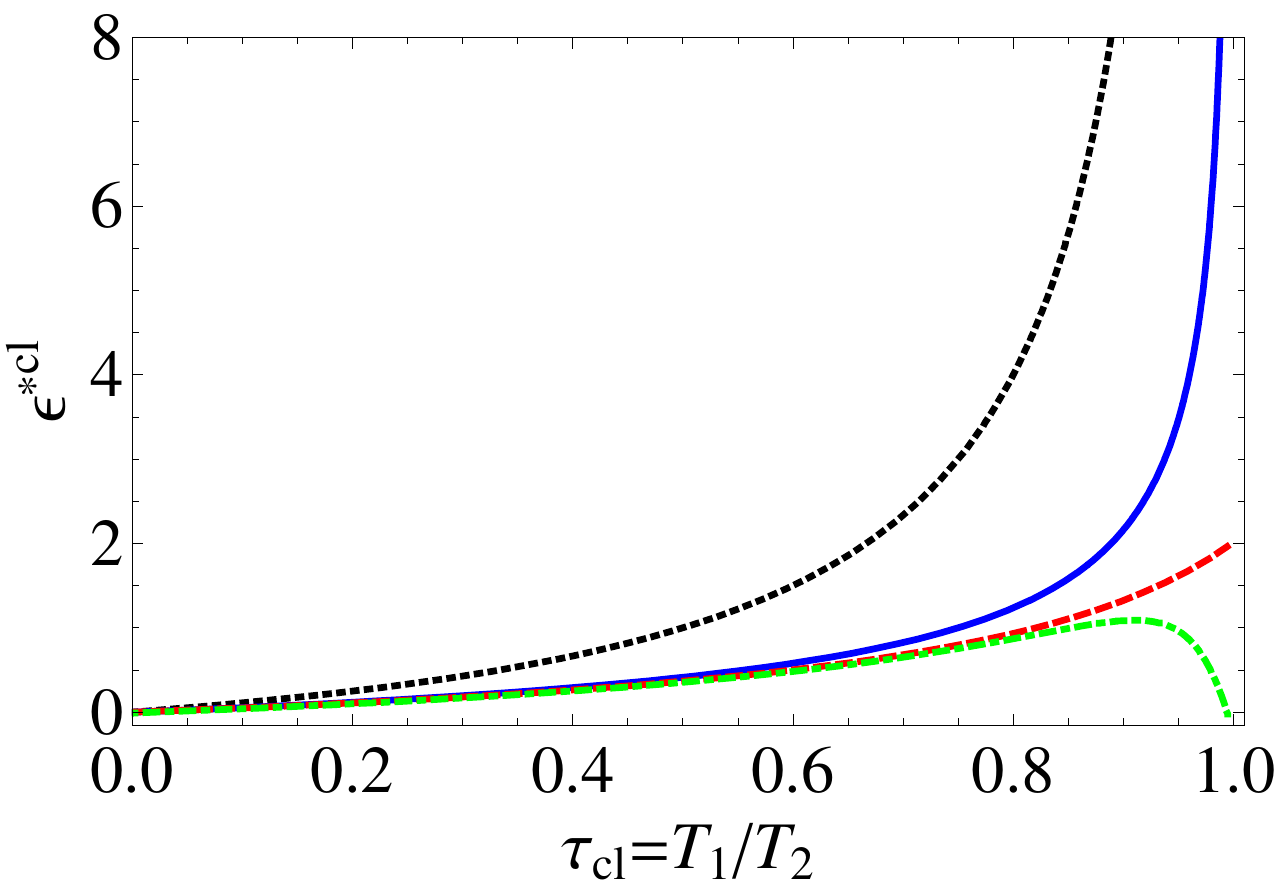}
\caption{Classical coefficient of performance at maximum figure of merit $\epsilon^{*cl}$ for the Otto refrigerator as a function of the temperature ratio $\tau_{cl}=T_1/T_2$. The blue (solid) line shows the adiabatic case (9), while the red (dashed) line and the green (dashed-dotted) lines show respectively the exact numerical result and the approximation (14) for the nonadiabatic  frequency modulation for $y = 0.01$. The black (dotted) line is the Carnot coefficient of performance $\epsilon_c= (\tau_{cl}^{-1}-1)^{-1}$.
\label{COP}}
\end{figure}

\textit{Adiabatic frequency modulation.}
We begin by considering quasistatic expansion/compression  characterized by $Q_{1,2}^\ast = 1$. We assume that the temperature of the hot reservoir obeys $\beta_2\hbar\omega_2 \ll 1$. We then have $\coth \beta_2\hbar\omega_2\simeq 1/(\beta_2\hbar\omega_2)$ and the figure of merit (7) simplifies to,
\begin{equation}
\label{8}
\chi_{ad} = \left(\frac{\omega_1}{\omega_2-\omega_1}\right) \left(\frac{\hbar \omega_1}{2}\coth\left(\frac{\beta_1 \hbar \omega_1}{2}\right) -\frac{\omega_1}{\beta_2\omega_2}\right).
\end{equation}
This expression  holds for arbitrary cold reservoir temperatures. We optimize eq.~\eqref{8} with respect to the  final frequency $\omega_2$, assuming, as commonly done, that all other parameters such as temperatures, cycle time and initial frequency $\omega_1$ are fixed. By solving  the equation $\partial \chi_{ad}/\partial \omega_2=0$,  we find that the maximum figure of merit is obtained when the frequencies satisfy, $\omega_1/\omega_2 = 1 -\sqrt{1-\tau}$, where  $\tau = \beta_2\hbar\omega_1 \coth(\beta_1\hbar\omega_1/2)/2$.  The  coefficient of performance at maximum $\chi$ in the  adiabatic limit follows as, 
\begin{equation}
\epsilon^\ast_{ad} = \frac{1}{\sqrt{1-\tau}}-1.
\label{h2}
\end{equation}
In the high-temperature regime, $\beta_i\hbar\omega_i \ll 1$, we have $\tau = \tau_{cl}=\beta_2/\beta_1$ and we  recover the known coefficient  of performance at maximum figure of merit, $\epsilon^{\ast cl}_{ad} = \sqrt{1 + \epsilon_\mathrm{c}} -1$,  of a classical refrigerator as derived in Refs.~\cite{yan84,yan90,vel97}. 
On the other hand, in the quantum limit where the cold reservoir is at  low temperature, $\beta_1 \hbar\omega_1 \gg 1$, we may
 use the expansion $\coth(\beta_1\hbar\omega_1/2) \simeq 1 + 2\exp(-\beta_1\hbar\omega_1)$ in eq.~(9)  to write  $\tau = \tau_q\simeq\beta_2\hbar\omega_1/2+ \beta_2\hbar\omega_1 \exp(-\beta_1\hbar\omega_1)$. The expression  $\epsilon^{\ast q}_{ad} = 1/\sqrt{1-\tau_q}-1 $ is the quantum  generalization of the Yan-Chen optimal coefficient of performance. The parameter $\tau $ has a simple physical interpretation as the ratio of the mean energies of the harmonic oscillator coupled respectively to the cold and hot reservoirs.

\begin{figure}[t]
\includegraphics[width=0.95\columnwidth]{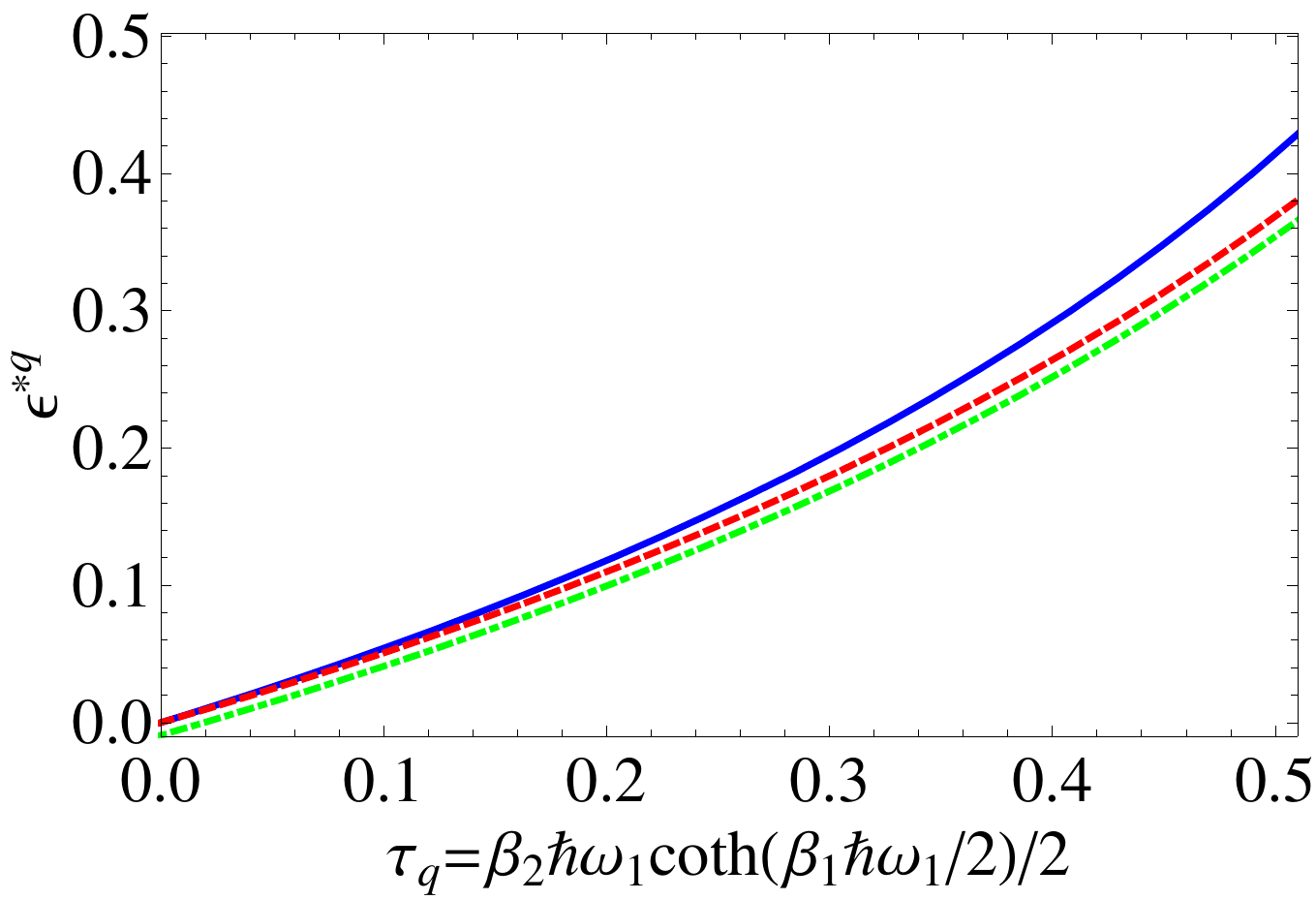}
\caption{Quantum coefficient of performance at maximum figure of merit $\epsilon^{*q}$ for the Otto refrigerator as a function of the energy ratio $\tau_q = \beta_2\hbar\omega_1 \coth(\beta_1\hbar\omega_1/2)/2$. The blue (solid) line shows the adiabatic case (9), while the red (dashed) line and the green (dashed-dotted) lines show respectively the exact numerical result and the approximation (14) for the nonadiabatic  frequency modulation for $y = 0.01$. 
\label{t_critical}}
\end{figure}

\textit{Nonadiabatic frequency modulation.}
For fast expansion/compression the  parameter $Q^\ast_{1,2} > 1$. We first treat the case of weakly nonadiabatic driving ($Q^\ast_{1,2}$ close to one) and defer the discussion of strongly nonadiabatic frequency modulation ($Q^\ast_{1,2}$ much larger than one) to the next section. In the limit of moderately fast evolution, we may linearize an  arbitrary frequency driving protocol as,
\begin{equation}
\omega_t = \omega_1 + (\omega_2 -\omega_1) t/t_0 = \omega_1 + \alpha t,
\end{equation}
with $\alpha = (\omega_2-\omega_1) /t_0$. The driving time $t_0$ is  here large, but finite, while it is infinitely large in the quasistatic limit. The parameter $Q^*_{1,2}$ may be calculated using the formulas given in Refs.~\cite{hus53,def08,def10} and expressed in terms of parabolic cylinder functions. We obtain, to lowest order in $\alpha$,
\begin{equation} 
Q^\ast_{1,2} = 1 + y,
\end{equation}
with $y = \alpha^2/(8 \omega_2^4)$. The  figure of merit (7) takes accordingly the form (for $\beta_2\hbar\omega_2 \ll 1$),
\begin{equation}
\chi_{na}\! = \!\frac{\left[\frac{\hbar\omega_1}{2} \text{c}\!\left(\frac{\beta_1\hbar\omega_1}{2}\right) - \frac{\omega_1}{\beta_2 \omega_2} (1+y)\right]^2}{\frac{\hbar\omega_1}{2}\!\left[\frac{\omega_2}{\omega_1}(1+y) -1\right]\text{c}\!\left(\frac{\beta_1\hbar\omega_1}{2}\right) + \frac{1}{\beta_2}\!\left[\frac{\omega_1}{\omega_2}(1+y)-1\right]}.
\end{equation}
Since the nonadiabatic correction $y$ is small, its $\omega_2$-dependence may be neglected. Maximizing eq.~(12) with respect to $\omega_2$ and keeping all  other parameters constant as before, we obtain the optimality equation, 
\begin{equation}
\left(\frac{\omega_1}{\omega_2}\right)^3 -\frac{2+\tau}{1+y}\left(\frac{\omega_1}{\omega_2}\right)^2 + 3\tau \left(\frac{\omega_1}{\omega_2}\right) - \frac{\tau^2}{1+y} = 0.
\end{equation}
Equation (13) is of third order in $\omega_2^2$ with two complex solutions (which are not relevant for the optimization problem) and one real solution which may  be evaluated analytically. We find, to lowest order in $y$,
\begin{equation}
\epsilon_{na}^{*} = \epsilon_{ad}^{*} - \frac{\left(2\, \tau + \sqrt{1-\tau}\right)}{(\tau -1)^2} y.
\label{16}
\end{equation} 
Equation (14) indicates that the nonadiabatic optimal coefficient of performance is  always smaller than the adiabatic one, as expected. The respective classical and quantum limits of $\epsilon_{na}^{*}$ are obtained replacing $\tau$ by $\tau_{cl}$ and $\tau_q$.

Figure 2 shows the adiabatic  optimal coefficient of performance $\epsilon^{*cl}$ (9) (blue solid line) in the classical regime, as well as  the nonadiabatic result obtained from the exact solution of eq.~(13) (red dashed line) and the approximation (14) to lowest order in $y$  (green dotted-dashed line). We observe that adiabatic and nonadiabatic coefficients of performance almost coincide for large temperature differences (small $\tau_{cl}$). However, they deviate significantly for small temperature differences  (large $\tau_{cl}$). Both are always smaller than the maximum Carnot formula (black dotted line). Figure 3 displays the corresponding results for the quantum optimal coefficient of performance $\epsilon^{*q}$ as a function of $\tau_q$, with a similar behavior. Quite generically, the amount of heat  $\langle Q_4\rangle $ that is extracted from the cold reservoir decreases with increasing temperature difference, or equivalently, decreasing cold temperature $T_1$ for fixed $T_2$.

\begin{figure}[t]
\begin{center}
\includegraphics[width=0.975\columnwidth]{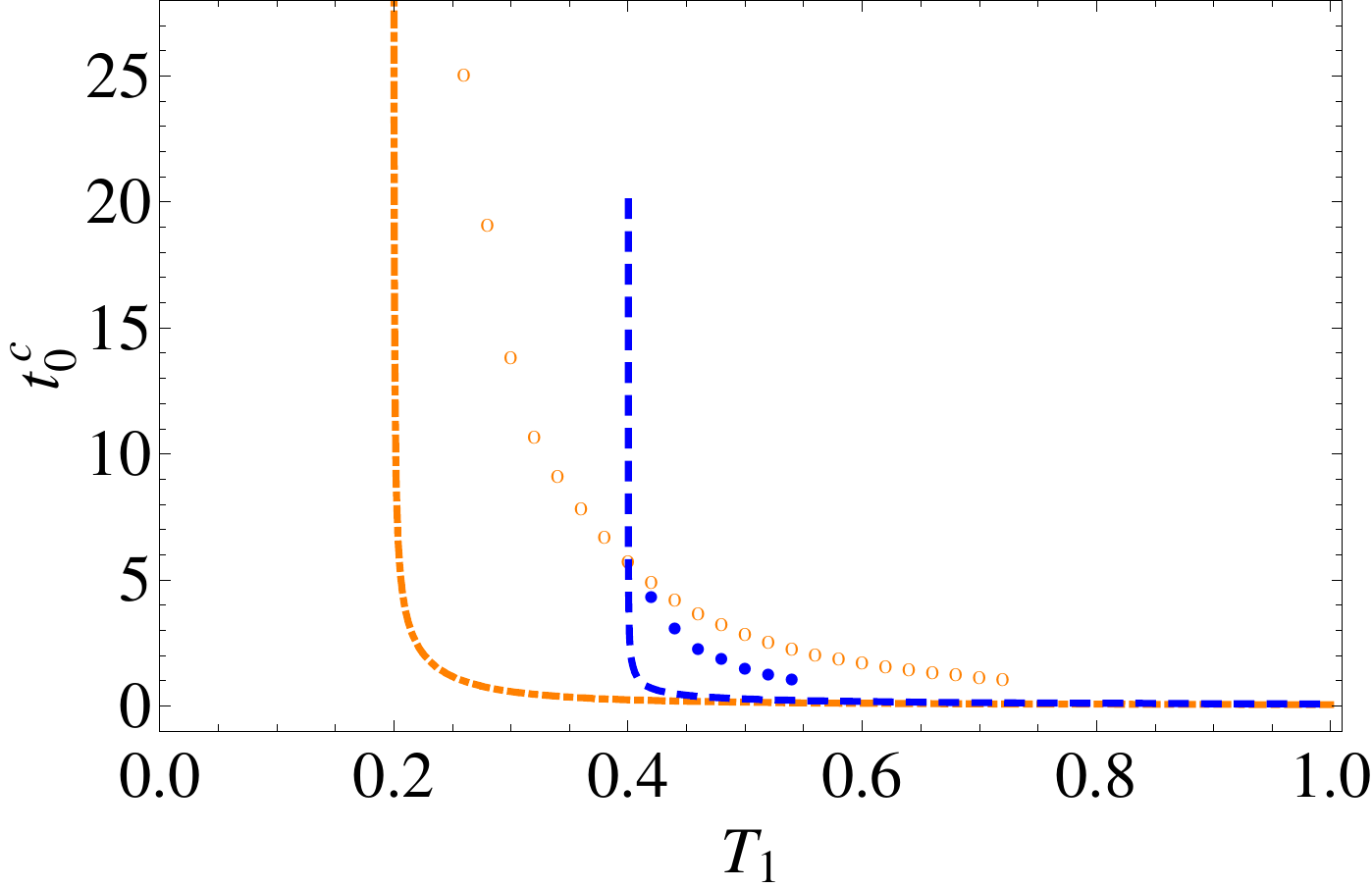}
\caption{Minimal driving time $t_0^c$  as a function of cold reservoir  temperature $T_1$. The blue dots and orange circles are  the exact numerical results obtained from eq.~(15), for $\omega_2=5$ and $\omega_2=10$, respectively. The blue (dashed) and orange (dotted-dashed) lines are the corresponding approximations (16). Parameters are $\hbar = 1,\, \omega_1 =1,\, \beta_2 =1$.
\label{fig4}}
\end{center}
\end{figure}
 
\section{Strongly nonadiabatic driving}
The parameter $Q^\ast_{1,2}$ increases with the degree of nonadiabaticity of the frequency protocols. Equation (4) reveals that the heat flow from the cold reservoir changes sign when $Q^\ast_{2}$ is larger than a given threshold. In this regime, the refrigerator stops cooling: the device consumes work to pump heat into both reservoirs, as noted in Ref.~\cite{rez11}. The physical origin of the breakdown of the cooling cycle is readily identified. Fast expansion/compression lead to nonadiabatic excitations of the harmonic oscillator that increase its mean energy. This extra energy is absorbed by the reservoirs during the thermalization phases. If the amount of energy induced by nonadiabatic transitions during the expansion step 3 exceeds the heat that is adiabatically extracted from the cold reservoir by the machine, the direction of the heat flow is reversed. This general mechanism explains why the cooling performance of the refrigerator is reduced by nonadiabatic frequency modulation,  until it comes to an end. The maximal value of  the parameter $Q^\ast_{2}$ that is allowed before  heat reversal occurs is given by,
\begin{equation}
Q^{\ast c}_{2} = \frac{\coth(\beta_1\hbar\omega_1/2)}{\coth(\beta_2\hbar\omega_2/2)}.
\end{equation}
Determining the corresponding minimal driving time $t_0^c$ is a difficult task as i) the latter  depends on the particular frequency protocol and ii) $Q^{\ast c}_{2}$ is a nontrivial function of $t_0^c$. In order to get an analytical estimate, we use the linear approximation (11) and find, 
\begin{equation}
t_0^c = \sqrt{\frac{(\omega_2 - \omega_1)^2 \coth(\beta_2\hbar\omega_2/2)}{8\omega_2^4 (\coth(\beta_1\hbar\omega_1/2) - \coth(\beta_2\hbar\omega_2/2))}}. 
\end{equation}
Figure 4 shows the exact critical driving time $t_0^c$ numerically obtained from eq.~(15) (circles and dots) and the analytical approximation (16) (dotted-dashed and dashed lines) as a function of the cold reservoir temperature $T_1$, for the linear driving  (10). We observe good agreement both for high and low temperatures\footnote{We show in the Appendix that the lowest order expression (11) for the parameter $Q^\ast_{1,2}$ also provides good approximations of the critical driving time $t_0^c$ for nonlinear driving protocols.}. The critical time $t_0^c$ increases exponentially for decreasing $T_1$ and diverges when the condition $\omega_1/\omega_2= T_1/T_2$ is verified. This equality corresponds to the maximum Carnot coefficient of performance. Consequently, strongly nonadiabatic expansion/compression are only possible for high $T_1$ (small $t_0^c$). For decreasing temperature $T_1$ (larger $t_0^c$), permitted cooling protocols get more and more adiabatic, confirming the behavior seen in figs.~2 and 3 at small $\tau_{cl,q}$. This phenomenon may be understood by noting that the heat extracted from the cold reservoir strongly decreases with decreasing temperature $T_1$, thus only allowing close to  adiabatic processes to ensure positive heat extraction.

\section{Conclusions} 
We have studied the finite time performance of a quantum Otto refrigerator  with a working medium consisting of a time-dependent harmonic oscillator. We have derived analytical expressions for the  coefficient of performance at maximum  figure of merit for slow and fast expansion/compression, both in the high-temperature (classical) regime and in the low-temperature (quantum) limit. We have further discussed the breakdown of the cooling cycle for strongly nonadiabatic driving protocols and obtained estimates for the minimal driving time allowed. Our findings should be helpful  for the design of harmonic quantum refrigerators  that run in finite time, from nanomechanical to ion trap systems \cite{aba12,zha14,dec15}.

\section{Appendix}
We here show that the first order nonadiabatic expansion of the parameter $Q^\ast_{1,2}$ given in eq.~(11) also yields a good estimate of the critical driving time $t_0^c$ for nonlinear driving. We consider the frequency protocol,
\begin{equation}
\omega_t^2 = \omega_1^2 + (\omega_2^2 -\omega_1^2) t/t_0 = \omega_1^2 + \bar \alpha t
\end{equation}
with $\bar \alpha = (\omega_2^2-\omega_1^2) /t_0$. The parameter $Q^\ast_{1,2}$ may be expressed in terms of Airy functions \cite{def08,def10}. Equation (11) then reads $Q^\ast_{1,2}=1+\bar y$ with $\bar y= \bar \alpha^2/(32\omega_2^6)$ and the critical driving time  is found to be,
\begin{equation}
t_0^c = \sqrt{\frac{(\omega_2^2 - \omega_1^2)^2 \coth(\beta_2\hbar\omega_2/2)}{32 \omega_2^6 (\coth(\beta_1\hbar\omega_1/2) - \coth(\beta_2\hbar\omega_2/2))}}. 
\label{eq18}
\end{equation}

\begin{figure}[!t]
\begin{center}
\includegraphics[width=0.975\columnwidth]{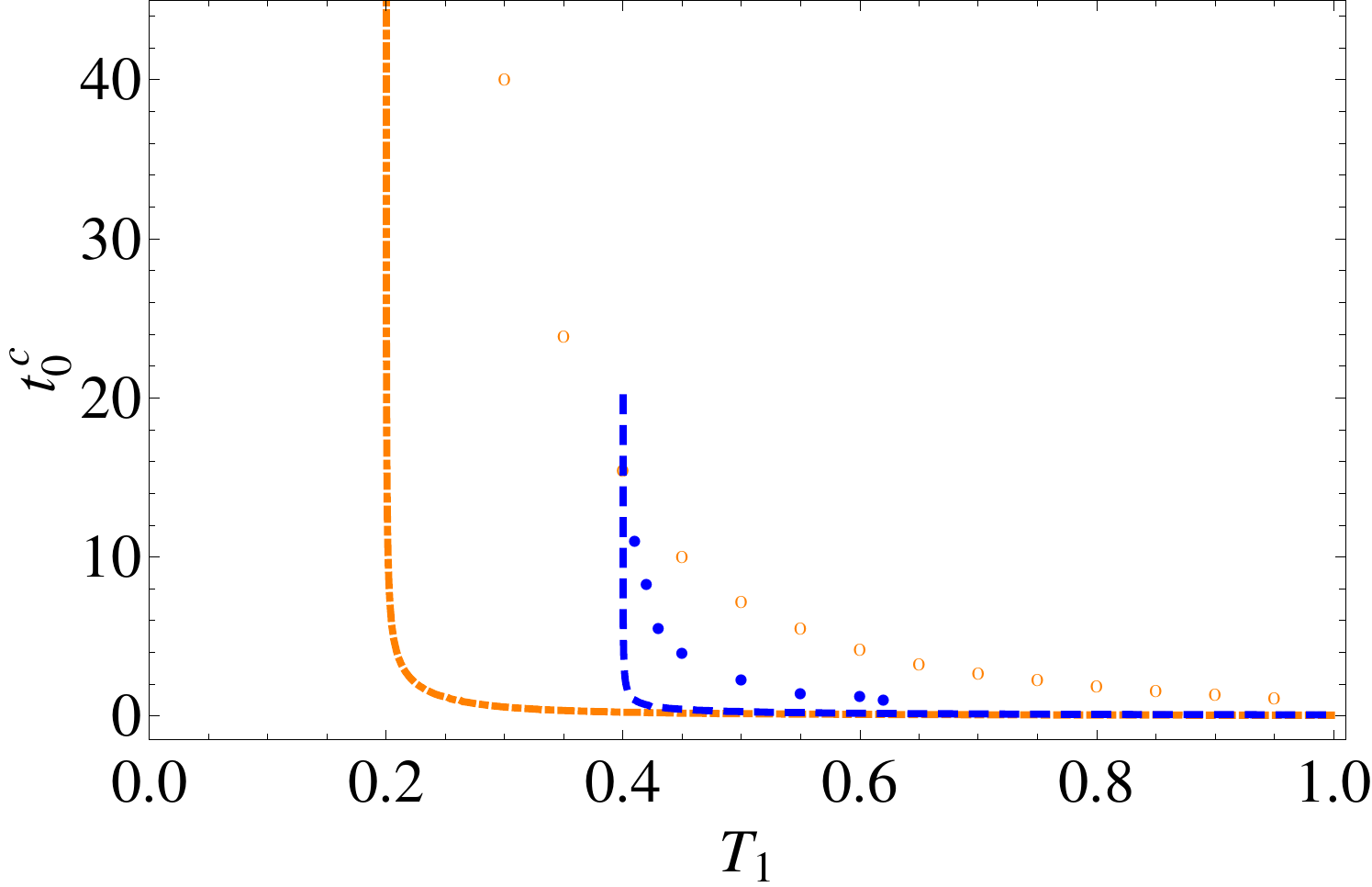}
\caption{Minimal driving time $t_0^c$  as a function of cold reservoir  temperature $T_1$ for the frequency driving protocol (17). The blue dots and orange circles are  the exact numerical results obtained from eq.~(15), for $\omega_2=5$ and $\omega_2=10$, respectively. The blue (dashed) and orange (dotted-dashed) lines are the corresponding approximations (18). Parameters are $\hbar = 1,\, \omega_1 =1,\, \beta_2 =1$.
\label{fig4}}
\end{center}
\end{figure}

Figure 5 shows the exact critical driving time $t_0^c$ numerically obtained from eq.~(15) (circles and dots) and the analytical approximation (18) (dotted-dashed and dashed lines) as a function of the cold reservoir temperature $T_1$, for the nonlinear driving  (18). We observe good agreement both for high and low temperatures as in fig.~4 for the linear frequency driving (10).

\acknowledgments
This work was partially supported by the EU Collaborative Project TherMiQ (Grant Agreement 618074) and the COST Action MP1209 "Thermodynamics in the quantum regime".

\end{document}